\documentclass[12pt]{article}
\usepackage{epsfig,cite}
\usepackage{amsmath}
\usepackage{amssymb}
\usepackage{amsfonts}
\usepackage{latexsym}
\usepackage{wasysym}
\usepackage{bm}
\def\bea{\begin{eqnarray}}
\def\eea{\end{eqnarray}}
\def\beq{\begin{equation}}
\def\eeq{\end{equation}}
\catcode`@=11 
\def\slash#1{\mathord{\mathpalette\c@ncel#1}}
 \def\c@ncel#1#2{\ooalign{$\hfil#1\mkern1mu/\hfil$\crcr$#1#2$}}
\def\lsim{\mathrel{\mathpalette\@versim<}}
\def\gsim{\mathrel{\mathpalette\@versim>}}
 \def\@versim#1#2{\lower0.2ex\vbox{\baselineskip\z@skip\lineskip\z@skip
       \lineskiplimit\z@\ialign{$\m@th#1\hfil##$\crcr#2\crcr\sim\crcr}}}
\catcode`@=12 
\def\twiddles#1{\mathrel{\mathop{\sim}\limits_
                        {\scriptscriptstyle {#1}}}}

\def\({\left(}
\def\){\right)}
\def\[{\left[}
\def\]{\right]}

\def\eps{\epsilon}

\def\a{\alpha_s}

\def\aq{\alpha_s(Q^2)}
\def\ab{\bar\alpha}

\def\ms{\overline{\rm MS}}
\def\as{\alpha_s}

\relax

\def\eg{{\it e.g.}}
\def\ie{{\it i.e.}}
\def\MS{\hbox{$\overline{\rm MS}$}}
\def\ms{{\overline{\rm MS}}}
\def    \hepph  #1 {{\tt hep-ph/#1}}
\def    \hepex  #1 {{\tt hep-ex/#1}}

\newsavebox\tmpfig

\begin{document}

\pagestyle{empty}

\begin{flushright}

IFUM-897-FT\\ GeF/TH/17-07\\
\end{flushright}

\begin{center}
\vspace*{0.5cm}
{\Large \bf A new prescription for soft gluon resummation}
 \\
\vspace*{1.5cm}
Riccardo Abbate,$^{a}$ Stefano~Forte$^{b}$ and Giovanni~Ridolfi$^{a}$
\\
\vspace{0.6cm}  {\it
{}$^a$Dipartimento di Fisica, Universit\`a di Genova and
INFN, Sezione di Genova,\\
Via Dodecaneso 33, I-16146 Genova, Italy\\ \medskip
{}$^b$Dipartimento di Fisica, Universit\`a di Milano and
INFN, Sezione di Milano,\\
Via Celoria 16, I-20133 Milano, Italy}\\
\vspace*{1.5cm}

{\bf Abstract}
\end{center}

\noindent
We present a new prescription for the resummation of the divergent
series of perturbative corrections, due to soft gluon emission,  to hard
processes near threshold in perturbative QCD (threshold
resummation). 
This prescription is based on Borel resummation, and 
contrary to  the commonly used minimal prescription, it does not 
introduce a dependence of resummed physical observables on
the kinematically unaccessible $x\to0$ region of parton
distributions. 
We compare results for resummed 
deep-inelastic scattering obtained using the Borel prescription and
the minimal prescription and exploit the comparison to
discuss the ambiguities related to the resummation procedure.

\vspace*{1cm}

\vfill
\noindent

\begin{flushleft} July 2007 \end{flushleft}
\eject

\setcounter{page}{1} \pagestyle{plain}
The resummation of logarithmically enhanced contributions to hard
processes near threshold~\cite{cnt,sterman}, 
such as deep-inelastic scattering and
Drell-Yan production at large values of the Bjorken $x$ variable (or
its analogue in the case of Drell-Yan), is
characterized by the fact that the effective scale of the process
is a soft scale related to the emission process. This means that for a
process with hard scale $Q^2$ the
resummation of large logs of $1-x$ effectively replaces the
perturbative coupling $\as(Q^2)$ with $\as(Q^2(1-x))$. In the space of
the variable $N$ which is conjugate to $x$ upon Mellin transformation,
where the resummation is more naturally performed, the effective
coupling is $\as(Q^2/N)$ and the soft limit $x\to1$ corresponds to $N\to\infty$.
 This result, which has
been understood long ago~\cite{bassetto} on the basis of an analysis
of evolution equations in the soft limit, and more recently in terms
of effective theories~\cite{efres}, is a simple consequence of the
fact~\cite{fr} that in the soft limit cross sections only depend on $x$
through the soft scale $Q^2(1-x)$, so this dependence can be
renormalization--group improved using standard techniques.

As the scale decreases, the strong coupling increases
and eventually it blows up at the Landau pole, so when
\beq \label{lpdef}
x=x_L\equiv 1-\frac{\Lambda^2}{Q^2}
\eeq
resummed results diverge, and physical observables can be determined
only
by specifying a prescription to treat this
divergence. A simple option, already discussed in ref.~\cite{bassetto},  is to perform the
resummation in $x$ space and cut off the phase space integration so
that the dangerous $x\geq x_L$ region is excluded. The option which is
more commonly used, however, is to perform the resummation in $N$
space, and reconstruct the result in $x$ space by Mellin inversion. In
this case, if the Mellin inversion is performed order by order in
perturbation theory, the series of resummed $x$-space contributions
diverges, and the problem is turned into that of summing a divergent
series~\cite{cmnt}. 

A commonly used way of treating this divergent series is the minimal
prescription (MP)~\cite{cmnt}, which, as we shall discuss in more
detail, 
is based
on the observation that the Mellin inversion integral of
the resummed $N$ space result exists if performed along a suitable
contour. Furthermore, the divergent series obtained from the
order-by-order Mellin inversion is an asymptotic expansion
of this integral. The minimal prescription, however, has the shortcoming that
upon convolution the partonic cross section does not vanish in the
unphysical $x>1$ region, which implies that physical
observables pick up a power-suppressed 
contribution from the unaccessible $x\to0$
region of parton distributions.

In ref.~\cite{frru} some of us suggested instead that the divergent
series could be summed using the Borel method, and showed how this
can be done at the leading logarithmic level for the logarithmic
derivative of the resummed partonic cross section.
Here we show how to perform this Borel resummation at any desired
logarithmic order for any physical observable (such as, say, the DIS
or Drell-Yan cross section at the hadronic level): namely, we give
here a
general Borel resummation prescription (BP). The availability
of several resummation prescriptions is per se useful as a way of
estimating the uncertainty of the resummation procedure. More
interestingly, we will show that the Borel prescription solves the
aforementioned problem of the minimal
prescription. Indeed, the BP leads to a resummed partonic cross
section which has the form of an $x$--space plus distribution,
such as found at finite perturbative order, and thus gives physical
observables by a convolution with parton
distributions in the standard way.
This is achieved  through the inclusion of a higher twist
term in the resummed result.

We will first summarize the properties of
the resummed result and in particular the divergence of the resummed
perturbative expansion. We will then describe the Borel resummation of
the resummed partonic cross section, and specifically discuss its
dependence on the choice of higher twist terms included in it. 
Finally we will compare the Borel prescription to the minimal
prescription, and in particular compare results for physical
observables obtained using either method.

In order to understand the origin of the divergence of resummed
results,  let us consider first as an example the computation of the
resummed leading log expression of 
\beq
\label{physandim}
\gamma(\as(Q^2),N)\equiv\frac{\partial\ln\hat\sigma(\frac{Q^2}{\mu^2},\as(\mu^2),N)}{\partial\ln Q^2},
\eeq
where $\hat\sigma\(\frac{Q^2}{\mu^2},\as(\mu^2),N\)$ is the Mellin transform
\beq
\label{mellin}
\hat\sigma\(\frac{Q^2}{\mu^2},\as(\mu^2),N\)=\int_0^1 \!dx\,x^{N-1}\bar\sigma\(\frac{Q^2}{\mu^2},\as(\mu^2),x\)
\eeq
of an observable $\bar\sigma\(\frac{Q^2}{\mu^2},\as(\mu^2),x\)$ such as the Drell-Yan
cross section or a deep-inelastic structure function, computed at the
parton level. In the soft limit, $\gamma(\as(Q^2),N)$
is computed up to terms which do not grow  as $N\to\infty$, and at 
the leading logarithmic level, it is a function of
$\alpha_s(Q^2)\ln\frac{1}{N} $ only. Explicitly,
\beq \label{LLgamma}
\gamma_{LL}(\aq,N)= g_1 \int_1^{N^a}\frac{dn}{n}\,\as(Q^2/n)
=-\frac{g_1}{\beta_0}\ln\[1+ \ab \ln\frac{1}{N}\].
\eeq
where $g_1$ is a constant, $a=1$ for deep--inelastic scattering and $a=2$ for Drell-Yan,
 we have used the explicit leading log form
of $\as(Q^2)$, 
\beq\label{loas}
\as(Q^2)=\frac{\as(\mu^2)}{1+\beta_0\as(\mu^2)\ln\frac{Q^2}{\mu^2}};
\qquad \beta_0=\frac{33-2n_f}{12\pi},
\eeq
and we have defined
\beq\label{abardef}
\quad \ab\equiv a\beta_0\aq.
\eeq

Clearly, $\gamma_{LL}(\as(Q^2),N)$ has a branch cut on the positive real axis
of the complex $N$ plane, starting at the Landau pole of $\alpha_s$ eq.~(\ref{loas}),
\beq
\label{lploc}
N_L=e^{\frac{1}{\ab}}.
\eeq
But if $\gamma_{LL}(\as(Q^2),N)$  were the Mellin transform of some
function $P_{LL}(\as(Q^2),x)$, it would be regular above some abscissa of
convergence $N_c$, \ie\ for all $\hbox{Re}(N)>N_c$. Hence,
$\gamma_{LL}(\as(Q^2),N)$ is not the Mellin transform of anything. However, to any
finite fixed perturbative order $M$ the inverse Mellin transform of
 $\gamma(\as(Q^2),N)$ is given by
\beq
\label{Pasympt}
P^{(M)}(\as(Q^2),x)=-\frac{g_1}{\beta_0}
\sum_{k=1}^M \frac{(-1)^{k+1}}{k} \ab^k\,
\frac{1}{2\pi i}\int_{\bar N-i\infty}^{\bar N+i\infty}dN\,
x^{-N}\ln^k\frac{1}{N};\qquad \bar N>0,
\eeq
where all Mellin inversion integrals can be computed exactly
[see the appendix, eq.~(\ref{exmel})]. It is easy to see that the limit of
$P^{(M)}(\as(Q^2),x)$ as $M\to\infty$ diverges. Indeed, if the limit existed,
then one could interchange the sum over $k$ and the integral over $N$,
but the sum over $k$ is then the Taylor expansion of
$\gamma_{LL}(\as(Q^2),N)$ eq.~(\ref{physandim}), which has finite
radius of convergence 
$|N|< N_L$, whereas the $N$ integral extends to infinity. 

In ref.~\cite{frru} we have computed the divergent series
eq.~(\ref{Pasympt}) explicitly and summed it {\it \`a la}  Borel. 
The approach of that reference however exploits the explicit
form of $\gamma_{LL}(\aq,N)$ eq.~(\ref{LLgamma}), and in particular
the fact that the integrand in eq.~(\ref{LLgamma}) has a simple
pole. We now present a generalization of that method which reduces to
it in the case of $\gamma_{LL}(\aq,N)$, but can be applied to any
resummed quantity.

We start with a Mellin-space resummed quantity $\Sigma$, function of
$\aq$, $\ln\frac{1}{N}$ and possibly
other kinematical variables such as the rapidity, which we will not
indicate explicitly. This resummed quantity is related to the partonic cross section
  $\hat\sigma\(\frac{Q^2}{\mu^2},\as(\mu^2),N\)$ eq.~(\ref{mellin}), or a
  quantity derived from it such as $\gamma$ eq.~(\ref{physandim}).
Now, in the soft limit $\hat\sigma$ eq.~(\ref{mellin}) can
be expanded  as~\cite{cmnt}
\beq\label{sigmexp}
\hat\sigma\(\frac{Q^2}{\mu^2},\as(\mu^2),N\)= \hat\sigma_0\exp\[\ln\frac{1}{N} g_1\(\ab
\ln\frac{1}{N}\)+g_2\(\ab
\ln\frac{1}{N}\)+\aq g_3\(\ab \ln\frac{1}{N}\)+\dots\],
\eeq
where $\hat\sigma_0$ is the  Born level result. 
It is thus convenient to expand the generic resummed quantity $\Sigma$ as 
\beq
\Sigma\(\aq,L\)=\lim_{M\to\infty}\sum_{k=1}^{M} h_{k}(\aq) L^{k}\label{2h1}
 \eeq
where we have defined 
\beq
\label{ldef}
L\equiv \ab\ln\frac{1}{N}=a\beta_0\aq\ln\frac{1}{N},
\eeq
with $\aq$ not necessarily
  given by its leading order expression. In the case of the
  computation of the partonic cross section, $\Sigma\(\aq,L\)$ is
  explicitly given by
\beq\label{restoobs}  
\hat\sigma\(\frac{Q^2}{\mu^2},\as(\mu^2),N\)=\hat\sigma_0\left[1+\Sigma\(\aq,L\)\right]. 
\eeq

For a generic resummed observable the series eq.~(\ref{2h1}) has
finite radius of convergence $|L|<1$ dictated by the location of the Landau
pole. Hence the term-by-term inverse Mellin
of the series eq.~(\ref{2h1}) is divergent. The divergent series can
be determined explicitly~\cite{frru} 
[see eq.~(\ref{exmel}) of the appendix] to compute the inverse
Mellin transform of $\Sigma\(\aq,L\)$ eq.~(\ref{2h1}), but with $M$ kept finite,
\beq\label{mellinv}
\bar\Sigma^{M}\(\aq,x\)\equiv \int_{\bar N-i\infty}^{\bar N+i\infty}\!\frac{dN}{2\pi i} \,x^{-N}\Sigma\(\aq,L\),\qquad\hbox{$M$ finite }.
\eeq
We get
\bea\label{ressig}
&&\bar\Sigma^M\(\aq,x\)=
\[\frac{R^M(x)}
{1-x}\]_++O\[(1-x)^0\],\\
\label{2Rdiv}
&&\> R^M(x)=\sum_{n=0}^{M}\Delta^{(n)}\(1\)\sum_{k=n}^{M}
\(\begin{array}{c}k\\n\end{array}\)c_{k}\ab^{k+1}
\ell^{k-n}+O\[(1-x)^0\],
\eea
where we have defined
\beq
\label{defpars}
c_k\equiv (k+1)h_{k+1}; \quad\ell\equiv\ln(1-x),
\eeq
$\Delta^{(n)}(z)$ is the $n$-th derivative of 
\beq\label{deldef}
\Delta(z)\equiv\frac{1}{\Gamma(z)},
\eeq
$O\[(1-x)^0\]$ denotes terms which are nonsingular in the limit
$x\to1$,
and for brevity we have
omitted the explicit dependence of the coefficients $h_k$ and of $R^M$
on  $\aq$. The divergent series which we wish to sum is then
\beq
\label{rlimdef}
 R(x)=\lim_{M\to\infty}  R^M(x).
\eeq

The divergence of $R(x)$ 
can be removed by performing a Borel transform with
respect to $\ab$, which gives 
\beq
\hat R(w,x)=\sum_{n=0}^{\infty}
\frac{\Delta^{(n)}\(1\)}{n!}\sum_{k=n}^{\infty}\frac{c_k}{\(k-n\)!}w^k
\ell^{k-n}.\label{boh11}
\eeq
The inner series has an infinite radius of convergence because its
coefficients are  factorially  smaller than those of the
series eq.~(\ref{2h1}). 
Because $\Delta(z)$ eq.~(\ref{deldef}) is an entire function of $z$,
it is easy to show that this implies that the outer series is also
convergent. Indeed, because the series $\sum_k c_k z^k$ is convergent
with the same radius as the series eq.~(\ref{2h1}), as $k\to\infty$ the coefficients
$c_k$ are bounded by some constant $K>0$, $|c_k|<K$. 
But this implies 
\beq\label{bound}
|\hat R(w,x)|\le K \sum_{n=0}^{\infty}
\frac{|\Delta^{(n)}\(1\)|}{n!}\sum_{k=n}^{\infty}\frac{1}{\(k-n\)!}|w^k
\ell^{k-n}|=K e^{|w\ell|}\sum_{n=0}^{\infty}
\frac{|\Delta^{(n)}\(1\)|}{n!}|w|^n
\eeq
which converges because of the absolute convergence of the power
series for $\Delta(z)$ eq.~(\ref{deldef}).

The original series can be recovered by inverting the Borel transform,
\beq
\label{invborel}
R(x)=\int_0^{\infty}\!dw\,e^{-\frac{w}{\ab}} \hat R(w,x),
\eeq
but 
the integral over $w$ in eq.~(\ref{invborel}) diverges at infinity:
 indeed, if we integrate the series term by term we recover
the original divergent series eq.~(\ref{2Rdiv}). We can cut
off the singularity by extending the integral only up to some upper bound
$C$. Because the series eq.~(\ref{boh11}) converges uniformly in the
interval $0\le w\le C$, we can integrate term by term, with the result
\bea
\label{1asympt}
R_B\(x,C\)&=& \int_0^{C}\!dw\,e^{-\frac{w}{\ab}}\hat R(w,x)
\\ 
&=&\sum_{n=0}^{\infty}\Delta^{(n)}\(1\)\sum_{k=n}^{\infty}\(\begin{array}{c}k\\n\end{array}\)c_k\frac{\gamma\(k+1,\frac{C}{\ab}\)}{k!}{\ab}^{k+1}\ell^{k-n}\label{2asympt},
\eea
where 
\beq\label{deftruncgam}
\gamma(k+1,z)\equiv\int_0^{z}\!dw\,e^{-w}w^k=k!\(1-e^{-z}
\sum_{n=0}^{k}\frac{z^n}{n!}\)
\eeq
is the truncated gamma function.
The series eq.~(\ref{2asympt}) for $R_B\(x,C\)$ 
has infinite radius of convergence, like that for $\hat R(w,x)$
eq.~(\ref{boh11}).

The Borel resummation of $\bar\Sigma\(\aq,x\)$ is obtained substituting
the expression for $R_B\(x,C\)$ eq.~(\ref{2asympt}) in
eq.~(\ref{ressig}).
Equation~(\ref{2asympt}) is not vey useful because it
requires the evaluation of a double series. However, we will now show
that the series can be summed through an integral representation which
is not harder to evaluate numerically than the minimal
prescription. Before doing this, let us discuss the properties of the
Borel resummation. 

First, it is easy to see that the divergent series we started from $R(x)$
eq.~(\ref{2Rdiv}) is an asymptotic expansion of its Borel resummation
$R_B\(x,C\)$ 
eq.~(\ref{2asympt}). To this purpose, we note that $R(x)$ and
$R_B\(x,C\)$ are related by
\beq\label{twistexp}
R_B\(x,C\)=R(x)-R_{ht}(x,C),
\eeq
where, using eq.~(\ref{deftruncgam}),
\beq
R_{ht}(x,C)=e^{-\frac{C}{\ab}}\sum_{n=0}^{\infty}\Delta^{(n)}\(1\)\sum_{k=n}^{\infty}\(\begin{array}{c}k\\n\end{array}\)c_k\ell^{k-n}\sum_{m=0}^{k}\frac{1}{m!}\(\frac{C}{\ab}\)^m\ab^{k+1}.\label{highertwist}
\eeq
Hence, $R_{ht}\sim e^{-\frac{1}{\ab}}$, so it vanishes
faster than any power of $\aq$ as $\aq\to0$. 
It follows that the difference between 
$R_B\(x,C\)$  and the sum of the first $N$ terms of $R\(x\)$ is of
order $\aq^{N+1}$, which proves that  $R\(x\)$ is an asymptotic
expansion of $R_B\(x,C\)$.

Furthermore, note that using the expression for $\aq$
\beq\label{llalpha}
\aq=\frac{1}{\beta_0\ln\frac{Q^2}{\Lambda^2}}\left[1+O(\aq)\right]
\eeq
we get
\beq\label{htres}
e^{-\frac{C}{\ab}}=\(\frac{\Lambda^2}{Q^2}\)^{C/a}\left[1+O(\aq)\right].
\eeq
This shows that cutting off the Borel inversion integral
eq.~(\ref{invborel}) at $w=C$  is equivalent to including a twist-$t$
contribution $R_{ht}(x,C)$, with 
\beq
\label{twist}
t=2+ \frac{2C}{a}.
\eeq
The divergence of the higher twist term then cancels that of the
divergent series, leading to a finite result. The value of  $C$ should
be  chosen in such
a way that no new, spurious higher twist terms are induced in physical observables.
The choice $C=a$ is minimal in that it corresponds to the inclusion of
a twist-four term, \ie\ a term of the first subleading twist.  

Let us now turn the Borel resummed expression $R_B\(x,C\)$
eq.~(\ref{2asympt}) into a more useful form.
In order to perform the sum over $n$, we introduce the Fourier
transform
\beq\label{truncfour}
\tilde{\Delta}_{\eta_0}\(\zeta\)\equiv \int_{-\infty}^{\infty}
\frac{d\eta}{2\pi}e^{- i\zeta\eta}\Delta\(1+\eta\)  \Theta(\eta+\eta_0)
\eeq
which satisfies
\beq
\label{invfour}
\int_{-\infty}^{\infty}
d\zeta e^{ i\eta\zeta}
\tilde{\Delta}_{\eta_0}\(\zeta\)=\Delta(1+\eta)\Theta(\eta+\eta_0),
\eeq
where $\Theta(\eta)$ is the Heaviside step function, and it is
necessary to introduce a cutoff at $\eta_0$ because 
the 
Fourier transform  of the function $\Delta (z)$ does not exist.
Rewriting  $R_B\(x,C\)$ eq.~(\ref{1asympt},\ref{boh11}) with
\beq\label{delder}
\Delta^{(n)}(1)= \int_{-\infty}^{\infty}\!d\zeta\,
\tilde{\Delta}_{\eta_0}\(\zeta\) \(i\zeta\)^{n}
\eeq
we can
perform the sum over $n$ explicitly:
\beq\label{ftasymptb}
R_B\(x,C\)
=\int_0^{C}\!dw\,e^{-\frac{w}{\ab}}\int_{-\infty}^{\infty}\!d\zeta\, \tilde{\Delta}_{\eta_0}
\(\zeta\)\sum_{k=0}^{\infty}\frac{c_k}{k!}\[w\(\ell+ i\zeta\)\]^k.
\eeq
where it  is sufficient to choose $\eta_0>0$ to ensure that the result is
independent of the choice of $\eta_0$.

\begin{figure}
\centering
\epsfig{width=.6\textwidth, file=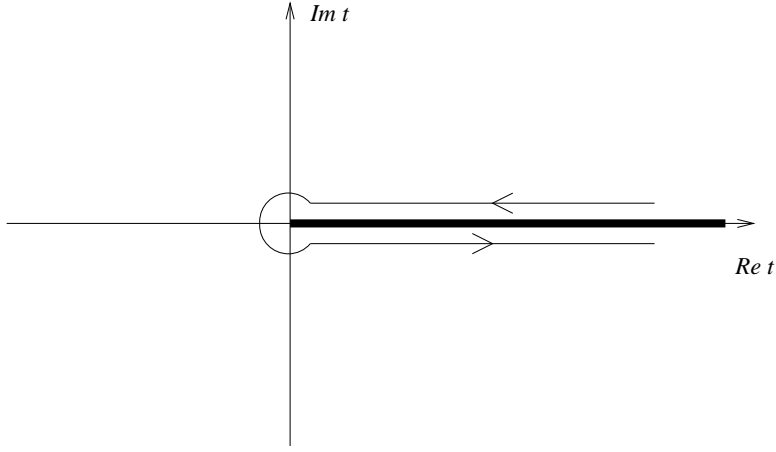}  
  \caption{The Hankel contour used in eq.~(\ref{2gammahankel}). \label{2Hankel}}
\end{figure}
Now, we observe that the sum over $k$ can be performed explicitly if the
factor of $k!$ in the denominator is removed. We do this as
follows. First, we write $1/k!=1/\Gamma(k+1)$ and we use the Hankel
representation of the Gamma function
\beq
\frac{1}{\Gamma\(z\)}=-\int_H\frac{dt}{2\pi i}e^{-t}\(-t\)^{-z}
\label{2gammahankel},
\eeq
where $H$ is the Hankel contour shown in fig.~\ref{2Hankel}. Furthermore,
because the integrand in eq.~(\ref{2gammahankel}) doesn't have any
other singularities in the complex plane besides the cut along the
positive real axis, the integral along the Hankel contour $H$ is equal
to the integral along the contour $H_1$
defined by 
\beq\label{newcont}
z_{H_1}=Re^{ i\theta},
\eeq
 with
$R\to\infty$ and $\eps\leq \theta\leq 2\pi-\eps$ with $\eps\to
0$. If we substitute in eq.~(\ref{ftasymptb}) the expression 
eq.~(\ref{2gammahankel}) with $z=k+1$ and the
integral over $t$ performed along $H_1$ we can integrate term by term 
over $t$ the sum over $k$, because
 the contour $H_1$ is always within the radius of
convergence of the series if $R$ eq.~(\ref{newcont}) is large
enough. We get
\bea
R_B\(x,C\)&=&\int_0^{C}\!dw\,e^{-\frac{w}{\ab}}\int_{-\infty}^{\infty}d\zeta
\tilde{\Delta}_{\eta_0}\(\zeta\)\int_{H_1} 
\frac{dt}{2\pi it}e^{-t} \sum_{k=0}^{\infty}c_k
\[-\frac{w}{t}\(\ell+ i\zeta\)\]^k\nonumber\\
&=&\int_0^{C}\!dw\,e^{-\frac{w}{\ab}}\int_{-\infty}^{\infty}d\zeta
\tilde{\Delta}_{\eta_0}\(\zeta\)\int_{H_1} \frac{dt}{2\pi it}
e^{-t} \Sigma'\(-\frac{w}{t}\(\ell+ i\zeta\) \),\label{2newborel}
\eea
where we have defined
\beq\label{sigprimdef}
\Sigma'(z)\equiv\frac{\partial}{\partial z}\Sigma(\aq,z)
\eeq
in terms of the function $\Sigma\(\aq,L\)$ eq.~(\ref{2h1}).

We can now remove the dependence of $\Sigma^\prime$ on $\zeta$ through the change of variables
\beq\label {xidef}
\xi=-\frac{t}{w\(l+ i\zeta\)} 
\eeq
whereby the contour $H_1$ eq.~(\ref{newcont}) 
is mapped onto a contour $\bar H_1$, which can be deformed back to
the contour $H_1$ for the new variable $\xi$.
The integral over $\zeta$ can
then  be performed using eq.~(\ref{invfour}) with the result
\beq
R_B\(x,C\)=\int_0^{C}\!dw\,e^{-\frac{w}{\ab}}\int_{H_1}
 \frac{d\xi}{2\pi i\xi} (1-x)^{w\xi}
 \Delta\(1+w\xi\)\Sigma'\(1/\xi\).\label{2borel}
\eeq    

The result eq.~(\ref{2borel}) can be already used as a resummation
prescription. However, it may be more convenient to rewrite it directly in terms of
the resummed observable $\Sigma$ rather than its partial
derivative. This is accomplished  integrating by parts:
\beq\label{3borel}
R_B\(x,C\)=
\int_0^C \!\frac{dw}{w}\,e^{-\frac{w}{\ab}}\int_{H_1}\frac{d\xi}{2\pi
  i}\frac{d}{d\xi}\bigg[w\xi e^{w\ell\xi}\Delta\(1+w\xi\)\bigg]\Sigma
\(1/\xi\),
\eeq
where the surface term vanishes provided only the radius $R$
eq.~(\ref{newcont}) of  the
contour in the $\xi$ plane is large enough, because $\Sigma\(1/\xi\)$
has a
discontinuity along the negative real $\xi$ axis that only extends from
the origin up to
the location of the Landau pole at $\xi=-1$. With straightforward manipulations we
can rewrite eq.~(\ref{3borel}) as
\beq\label{finborel}
R_B=\int_{H_1}\frac{d\xi}{2\pi i}
\bigg[W(C,\ell,\xi)+\frac{1}{\ab}
\int_0^C\!dw\,W(w,\ell,\xi) \bigg]\Sigma\(1/\xi\),
\eeq
where we have defined
\beq\label{wdef}
W(w,\ell,\xi)\equiv we^{-\frac{w}{\ab}(1-\ell\xi\ab)}\Delta(1+w\xi)=
we^{-\frac{w}{\ab}}
(1-x)^{w\xi}\Delta(1+w\xi).
\eeq

The Borel prescription for the resummation of the divergent series
eq.~(\ref{2Rdiv}) consists of taking
\beq
\label{borsigma}
\bar\Sigma\(\aq,x\)=\left[\frac{R_B(x, C)}{1-x}\right]_+,
\eeq
with $R_B(x, C)$ given by either of the equivalent expressions 
eq.~(\ref{2borel}) or eq.~(\ref{finborel}). The integrand of the $\xi$ integral
has a cut
along the negative real $\xi$  axis for $-1\le\xi\le0$, and it is regular
elsewhere; the closed  contour $H_1$ encircles this cut. The value of
$C$ is related by eq.~(\ref{twist}) to the twist of the contribution
which is included in order to  get a finite resummed result; the
minimal choice is $C=a$, corresponding to the inclusion of a twist
four term.

Let us now briefly discuss some properties of the Borel resummed
result eq.~(\ref{borsigma}) and then compare it to the result
obtained using the minimal prescription. 
First, let us determine it explicitly in the simplest case in which we
take as resummed observable
\beq\label{siggam}
\Sigma\(\aq,L\)=\gamma_{LL}(\aq,N),
\eeq
with $\gamma_{LL}(\aq,N)$    given by eq.~(\ref{LLgamma}). In this case,
it is convenient to use eq.~(\ref{sigprimdef}), since
\beq\label{parsigma}
\Sigma^\prime(1/\xi)=
-\frac{g_1}{\beta_0} \frac{\xi}{1+\xi}
\eeq
so the $\xi$ integral is straightforward:
\bea\label{padllres}
R_B\(\aq,x,C\)&=&-\frac{g_1}{\beta_0}\int_0^C\!
dw\,e^{-\frac{w\(1+\beta_0 \a l\)}{\beta_0 \a}}\Delta\(1-w\)\nonumber\\
&=&-\frac{g_1}{\beta_0}\int_0^C
\!dw\,\[\frac{\Lambda^2}{Q^2(1-x)}\]^w\Delta\(1-w\)
\eea
which coincides with the result of ref.~\cite{frru}.
The next-to-leading log result can be analogously determined in closed form.

The resummed result eq.~(\ref{borsigma})  has the form of a plus
distribution, whose action on any test function $f(x)$ leads to a
finite result provided the numerator $R_B(x,C)$ is integrable as a
function of $x$ between $0$ and $1$. However, the explicit result
eq.~(\ref{padllres}) suggests that this is the case only if $C$ is not
too large. Indeed, the integrand of eq.~(\ref{2borel}) is integrable
over $x$ as $x\to1$ only if 
\beq\label{ineq}
\hbox{Re}(w \xi)>-1.
\eeq
The path $H_1$ must
intersect the negative real axis at  some $\xi=\xi_0<-1$ because of the
cut up to the Landau pole at $\xi=-1$. Hence the condition becomes
$\hbox{Re}(w)<1$ which is
 violated whenever $C\ge1$. 

Nevertheless, for
all $C$ the
action of $\bar\Sigma\(\aq,x\)$ is well defined by analytic continuation, and this allows 
 its numerical implementation. Indeed, consider the action upon
 integration of $\bar\Sigma(\aq,x)$ on the test function 
\beq\label{testf}
\tau(x)=(1-x)^n.
\eeq
We get
\beq
\int_0^1\!dx\,\bar \Sigma\[\aq,x\]\tau(x)
=\frac{1}{2\pi i}\int_0^C \!dw\, e^{-\frac{w}{\ab}}
\oint\! \frac{d\xi}{\xi}\, \Sigma'(1/\xi)
\Delta(1+w\xi)\frac{1}{n+w\xi}.
\label{int}
\eeq
The integrand is regular at  $w=-n/\xi$, because, for any negative
integer $-n$,
\beq\label{delexp} 
\frac{\Delta(1+ z)}{(z+n)} =(-1)^{n-1}(n-1)!\[1+O(z+n)\],
\quad (n\ge0~\hbox{integer})
\eeq
hence the integral eq.~(\ref{int})
exists for all $C$.
 This immediately implies that $\bar\Sigma(\aq,x)$ 
is a distribution which gives finite results when integrated over 
any test function $\tau(x)$ which is analytic in the neighbourhood of 
$x=1$.

In practice, for numerical computations one can proceed as follows:
the quantity of interest is typically a convolution of
$\Sigma\(\aq,x\)$ with a parton density $q(x)$, of the form
\beq
\label{convph}
\int_x^1\frac{dz}{z}\bar\Sigma(\aq,z)q\left(\frac{x}{z}\right)
=\int_x^1 dz
\frac{R_B(z,C)}{1-z}
\left[\frac{1}{z}q\left(\frac{x}{z}\right)-q(x)\right]
-q(x)\int_0^xdz\frac{R_B(z,C)}{1-z}.
\eeq
The first term on the right-hand side of this equation only leads to a
convergent integral if $R_B(z,C)$ is integrable, which in turn
requires $C<1$ as discussed above. However, we can rewrite this
integral defining
\beq\label{gdef}
g(x,z)=\frac{1}{1-z}
\left[\frac{1}{z}q\left(\frac{x}{z}\right)-q(x)\right],
\eeq
as follows
\beq
\label{subtract}
\int_x^1 dz
\frac{R_B(z,C)}{1-z}
\left[\frac{1}{z}q\left(\frac{x}{z}\right)-q(x)\right]=
\int_x^1 dz\,
R_B(z,C)\left[g(x,z)-g(x,1)\right]+ g(x,1)\int_x^1 dz\,R_B(z,C).
\eeq
The second integral on the right-hand side of eq.~(\ref{subtract})
can be computed analytically using eq.~(\ref{int}) with $n=0$,
while the first integral is now convergent even if  $R_B(z,C)$ is not
integrable, provided only $R_B(z,C)\twiddles{z\to1}(1-z)^b$ with
$b>-2$, which now only requires $C<2$.
If $R_B(z,C)$ is even more divergent around $z=1$ one simply iterates
the procedure. This allows one to choose an arbitrarily large value of $C$.

We finally compare the Borel prescription eq.~(\ref{borsigma})
to the minimal prescription. Consider first what happens when we apply either
of them to a quantity whose inverse Mellin transform does exist, such
as $\Sigma(\aq,L)$ eq.~(\ref{2h1}) when  $M$ is kept finite. In such
case, the minimal prescription simply gives this inverse Mellin
transform. Taking for example eq.~(\ref{2h1}) with $M=1$, $h_1=1$, \ie, 
$\Sigma(\aq,L)=L$, the minimal prescription gives [see appendix, eq.~(\ref{exmel})]
\beq\label{exmp}
\bar\Sigma^{1,\,{\rm
  MP}}(\aq,x)=\ab\(\[\frac{1}{\ln\frac{1}{x}}\]_++\delta(1-x)\).
\eeq
If instead we apply the Borel prescription, we get a result that
differs from the inverse Mellin first, because terms which are either finite or zero
as $x\to1$ are neglected, and furthermore, because the higher twist
correction eq.~(\ref{highertwist}) is included. In the previous
example, this gives, instead of eq.~(\ref{exmp}),
\beq\label{exbp}
\bar\Sigma^{1,\,{\rm
  BP}}(\aq,x)=\ab\[\frac{1}{1-x}\]_+\(1-e^{-\frac{1}{\ab}}\).
\eeq

If one applies the MP to a function $\Sigma(\aq,L)$ whose Mellin
transform does not exist because of a branch cut from $N_L$
eq.~(\ref{lploc}), such as a typical resummed quantity, the ensuing
$x$--space result $\bar\Sigma^{\rm MP}(\aq,x)$ does not vanish in the unphysical region $x\ge 1$.  
It follows that a
physical observable $\sigma^{\rm MP}(x)$,
 computed combining a partonic cross section $\hat\sigma^{\rm MP}(N)$
 eq.~(\ref{sigmexp}) with a parton distribution $q(N)$ [with inverse
 Mellin $\bar q(x)$], has the form
\beq\label{mpobs}
\sigma^{\rm MP}(x)=\int_0^1\!\frac{dy}{y}\, \bar\sigma^{\rm
  MP}\(\frac{x}{y}\) \bar q(y),
\eeq
{\it i.e.} it receives a contribution from the 
unphysical $0\le y\le x$ region of parton densities (see appendix B of 
ref.~\cite{cmnt}), though it
has been shown in ref.~\cite{cmnt} that this contribution is power 
suppressed. Furthermore, there are practical
difficulties in the construction of the $x$--space result
$\bar\Sigma^{\rm MP}(\aq,x)$ which is needed \eg\ if one wants to use 
a resummed result with $x$ space parton distributions, related to the fact 
that  the MP result for $\bar\Sigma^{\rm
  MP}(\aq,x)$ displays an oscillatory behaviour~\cite{cmnt}.

The main advantage of the Borel
prescription result eq.~(\ref{borsigma}) is that it gives  directly
$\bar\Sigma^{\rm BP}(\aq,x)$ in $x$ space, in the form of a plus distribution 
as those found order by order in perturbation theory. Physical
observables are obtained from it by standard convolution with parton
distributions
in the physical region:
\beq\label{bpobs}
\sigma^{\rm BP}(x)=\int_x^1\!\frac{dy}{y}\, \bar\sigma^{\rm
  BP}\(\frac{x}{y}\) \bar q(y).
\eeq
This is accomplished by including power suppressed terms
order by order in the physical region,  as explicitly shown in
eq.~(\ref{exbp}). As already mentioned, it is convenient to choose 
$C$ in such a way that
these power suppressed terms combine with those which already appear
at higher orders in the Wilson expansion. In fact, it has been argued
in ref.~\cite{gardi} that in the large $x$ limit  the dominant higher
twist contributions are those which mix upon renormalization with the
leading twist. Be that as it may, with the minimal choice $C=a$
eq.~(\ref{twist})
the
ambiguity introduced by the BP may be cancelled by an equal and
opposite ambiguity from a conventional higher twist term, as already
discussed in ref.~\cite{frru}.

A further advantage of the BP is that the
non-logarithmically enhanced  terms which are generated by the exact
Mellin inversion of $\Sigma(\aq,L)$  can be included or excluded at will. 
Indeed, the computation of the exact Mellin inverse, as done in the MP, is
not necessarily 
advantageous if the resummed $\Sigma(\aq,L)$ is only computed in
the large $N$ limit to begin with. For instance, in the simple example
considered above, the 
series of terms generated by
the expansion of $1/\ln\frac{1}{x}=1/(1-x)-1/2-(1-x)/12+\dots$ in eq.~(\ref{exmp})  does
not necessarily 
provide a better approximation to the exact $O(\as)$ expression of
$\Sigma$ than the purely logarithmic contribution $1/(1-x)$ 
included in the BP result eq.~(\ref{exbp}). This is to be contrasted with the
case of
specific classes of non-enhanced~\cite{lor} or even
suppressed~\cite{akh} terms whose resummation might be advantageous. 
Now, in the BP it is possible
to choose whether to perform the Mellin inversion exactly or in the
large $x$ limit, unlike in the MP where the inversion is always
performed exactly. Indeed,
 it is easy to modify the BP  in such a way that
when applied to $\Sigma(\aq,L)$ eq.~(\ref{2h1}) with finite $M$
it coincides with its exact inverse Mellin up to higher twist
terms. For this, it is sufficient to use the method described in the
appendix to determine the Mellin inversion eq.~(\ref{exmel})
exactly. In practice, it is sufficient to replace everywhere $(1-x)$
with $\ln \frac{1}{x}$ in
eqs.~(\ref{ressig}-\ref{2Rdiv}) and in the final results eq.~(\ref{2borel})
or eq.~(\ref{wdef}).  

The ambiguity in the resummation procedure can be estimated by
comparing results obtained using the Borel and minimal
prescriptions.  
In order for the comparison to be significant, we must
compare the convolution of the result with a test parton distribution:
indeed, the resummed partonic quantity 
$\bar\Sigma(\aq,x)$ is a distribution, rather than a function
proper. Furthermore, the different treatment of non logarithmically
enhanced terms between BP and MP is only allowed in the region where
the resummed logs are large: indeed, away from that region
any resummed result must reduce to the fixed
order. Hence, we must compare results matched to the fixed
order.

To this purpose, we have determined a matched  result for a physical
observable using the 
MP and BP. We consider the quark coefficient function for the
deep--inelastic structure function $F_2$ in the \MS\ scheme,
$C_{2,\,q}^{\ms}\(Q^2/\mu^2,\alpha(\mu^2),N\)$. We then determine the
resummed expression for this quantity with $\mu^2=Q^2$, up to the
next-to-leading log level, \ie\  we use eq.~(\ref{sigmexp}) for
$\sigma\(1,\as(Q^2),N\)$, including the
contributions $g_1$ and $g_2$, as given \eg\ in
ref.~\cite{vogt}. This is our resummed observable $\Sigma(\aq,L)$, to
be matched to the standard $O(\as)$ result for
$C_{2,\,q}^{\ms}\(1,\alpha(Q^2),N\)$~\cite{buras}. 
We further take a model quark distribution given by
\beq\label{qpdf}
\bar q(x)=x^{-1/2}(1-x)^3;\quad q(N)= \frac{\Gamma(4)
  \Gamma\(N-\frac{1}{2}\)}{\Gamma\(N+\frac{7}{2}\)} .
\eeq

\begin{figure}[t]
\begin{center}
\epsfig{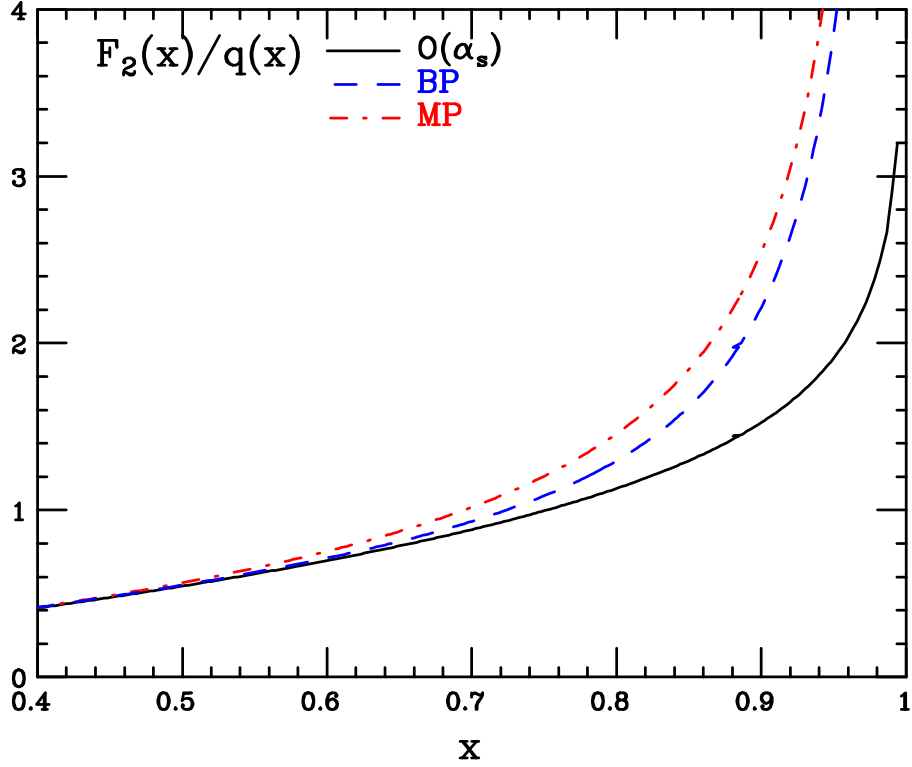}  
\end{center}
\begin{center} 
 \caption{Resummation of the 
quark coefficient function $C_{2,\,q}^{\ms}\(Q^2/\mu^2,\alpha(\mu^2),N\)$
for 
    the deep-inelastic structure function $F_2$. The resummation is
    performed up to the next-to-leading logarithmic level  matched
    to the $O(\aq)$ fixed order result. 
We plot as a function of $x$ the
    structure function normalized to the parton distribution. 
The three curves are, top to bottom,
    the minimal prescription eq.~(\ref{mpres}), the Borel prescription 
eq.~(\ref{bpres}) and the fixed $O(\aq)$ result. We take $\aq=0.2$.
 \label{resplot}}\end{center}
\end{figure}
The minimal prescription is then constructed by computing
\beq
\label{mpres}
F_{2,q}^{\rm MP}(Q^2,x)=\int_{C_{\rm MP}} \frac{dN}{2\pi i} x^{-N}
\[C_{2,\,q}^{\ms}\(1,\alpha(Q^2),N\)+\Sigma(\aq,L)-\Sigma^{\rm NLO}(\aq,L)
\]q(N),
\eeq
where $C_{\rm MP}$ is the standard minimal prescription
contour~\cite{cmnt}, $\Sigma(\aq,L)$ is the resummed coefficient discussed above, and 
$\Sigma^{\rm NLO}(\aq,L)$ is its expansion up to order $\aq$, namely
\beq
\label{nsubtr}
\Sigma(\aq,L)-\Sigma^{\rm NLO}(\aq,L)=O\(\alpha^2(Q^2)\).
\eeq
The Borel prescription is constructed computing 
\beq
\label{bpres}
F_{2,q}^{\rm BP}(Q^2,x)=\int_x^1 \frac{dy}{y}
\[\bar C_{2,\,q}^{\ms}\(1,\alpha(Q^2),y\)+\bar \Sigma(\aq,y)-\bar\Sigma^{\rm NLO}(\aq,y)
\]\bar q\(\frac{x}{y}\),
\eeq
where $\bar C_{2,\,q}^{\ms}\(1,\alpha(Q^2),x\)$ is the inverse Mellin
transform of $C_{2,\,q}^{\ms}\(1,\alpha(Q^2),N\)$, $\bar
\Sigma(\aq,x)$ is constructed from $\Sigma(\aq,L)$ using eq.~(\ref{borsigma}) 
with $R_B(x,C)$ given by eq.~(\ref{finborel}) and (for DIS) $a=1$. We
take $C=1$, which corresponds to the inclusion of a twist-four term; the
convolution integral in eq.~(\ref{bpres}) can then be computed with  
one subtraction eq.~(\ref{subtract}). Finally, 
$\bar\Sigma^{\rm NLO}(\aq,x)$ is the  expansion of $\bar
\Sigma(\aq,x)$ up to order $\aq$,
\beq
\label{xsubtr}
\bar\Sigma(\aq,x)-\bar\Sigma^{\rm NLO}(\aq,x)=O\(\alpha^2(Q^2)\).
\eeq

\begin{figure}[t]
\begin{center}
\epsfig{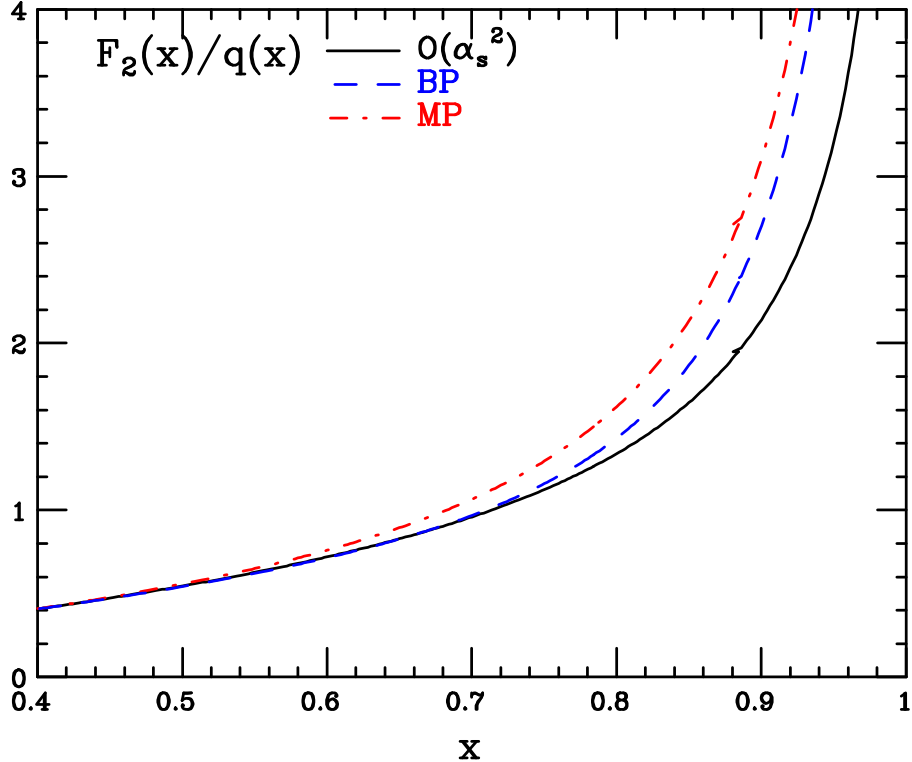}  
\end{center}
\begin{center} 
 \caption{Same as figure 2, but  with the resummation
    performed up to the next-to-next-to-leading logarithmic level  and
    matched
    to the $O(\alpha_s^2)$ fixed order result. 
 \label{resplot2}}\end{center}
\end{figure}
The results obtained using the MP and the BP are compared to each
other and to the fixed $O(\aq)$ result in Fig.~\ref{resplot}. The
structure function eqs.~(\ref{mpres}),(\ref{bpres}) is plotted as a
function of $x$, normalized to the parton distribution
eq.~(\ref{qpdf}): namely, we plot $\frac{F_{2,q}(Q^2,x)}{\bar q(x)}$.  We
take $\aq=0.2$. Note that $\bar q(x)$
vanishes very rapidly as $x\to1$. The comparison shows that the effect
of the resummation is sizable for $x\gsim 0.6$ and becomes of order $100\%$
when $x\gsim 0.8$, where however $F_2(x)$ is very
small. Interestingly, while the percentage difference between the MP
and BP tends to zero both as $x\to1$ and $x\to0$, in the intermediate
region $0.6\lsim x\lsim 0.8$ where the resummation is important the
two prescriptions lead to rather different results. 

We have checked that the replacement of $(1-x)\to\ln\frac{1}{x}$ in the BP 
has a negligible effect. This ensures that the difference between the 
MP and BP is not due to a different treatment of non-logarithmically 
enhanced terms. Furthermore, we have verified that increasing the value
of $C$ eq.~(\ref{1asympt}) from one to  1.8 also has essentially no effect.
This agrees with expectations based on the results of ref.~\cite{frru}:
there, it was found that $R_B(x,C)$ is stable upon variations of $C$  
unless $x>x_L$ eq.~(\ref{lpdef}), so the same should hold for 
physical observables where the region  of very large $x$ shouldn't
weigh too much. 
Finally, in Fig.~\ref{resplot2} we repeat the same calculation but
adding an extra logarithmic order in the resummed result and matching 
to the fixed order result computed at $O(\as^2)$. The difference
between the fixed--order and resummed results is now smaller, as it
ought to be, but the difference between MP and BP has not decreased,
thereby showing that this difference is not compensated by the
inclusion of higher logarithmic orders. 

We must conclude that the difference between the MP and the BP indicates
that
the ambiguity in the resummation procedure is sizable: a fact which is
rather well known in the context of transverse momentum resummation
(see \eg\ ref.~\cite{stirling}), but not equally obvious for threshold
resummation.

In summary, we have presented a new prescription for the resummation of
the divergent
series of logarithmically enhanced terms which is obtained from
threshold resummation. The divergent series is summed through the
Borel method, and the divergence in the Borel inversion integral is
removed through the inclusion of a suitable higher twist term. This
term can be chosen to be of any twist, but the minimal choice is to
take it as a twist four contribution. We
have described the practical implementation of this prescription and
demonstrated its application to the threshold resummation of a
deep-inelastic coefficient function, which we have compared to the 
commonly used minimal prescription.

The Borel prescription and minimal prescription have somewhat
complementary advantages and disadvantages: the minimal prescription
is naturally implemented in $N$ space, so it is easy to use with
$N$--dependent parton distributions. However, in $x$ space the MP leads to
partonic cross sections which do not vanish in the unphysical $x>1$
region and its implementation is less straightforward. The Borel
prescription directly gives an $x$ space result which has the form of  a 
plus distribution such as found in fixed order perturbative computations. 
However, its $N$ space form can only be obtained by
performing the Mellin transform numerically, and its convolution
with a parton distribution must be determined by numerical
integration. The Borel prescription also has the advantage that it is
possible to control the inclusion of non-logarithmically enhanced
terms in the resummation, but it has the disadvantage that it requires
the inclusion of higher twist contributions.

Comparison of results obtained using the Borel prescription and the
minimal prescription suggests that the ambiguity in threshold
resummation is sizable. The extension of this method to the case of
resummation of transverse momentum distributions will be presented
elsewhere.
\medskip

\noindent{\bf Acknowledgement:} We thank Paolo Nason for illuminating
discussions.
S.F. acknowledges partial support from the  Marie Curie Research
Training Network HEPTOOLS under contract MRTN-CT-2006-035505.

\subsection*{Appendix}
In ref.~\cite{fr} we have determined the  Mellin
transform of any function of $\ln N$ to all orders in $\ln (1-x)$, up
to terms which vanish as $x\to 1$ as a power of $1-x$. However,
the exact Mellin transform  can also be computed~\cite{frru}.
Indeed, the standard Euler integral representation of the Gamma function
implies that
\beq
\label{eulerdef}
\int_0^1dx x^{N-1} \left[\ln^{\eta-1}\frac{1}{x}\right]_+
=\Gamma(\eta) \left(N^{-\eta }-1\right).
\eeq
so
\beq\label{eulerinv}
\frac{1}{2\pi i}\int_{\overline N- i\infty}^{\overline N+ i\infty} 
dNx^{-N}N^{-\eta}=\Delta(\eta)\[\ln^{ \eta-1}\frac{1}{x}\]_+ +\delta(1-x),
\eeq
where $\Delta(\eta)\equiv\frac{1}{\Gamma(\eta)}$.
It follows that the exact inverse Mellin transform of $\ln^k\frac{1}{N}$ is
\bea\label{exmel}
\frac{1}{2\pi  i}
\int_{\overline N- i\infty}^{\overline N+ i\infty} dNx^{-N}\ln^k\frac{1}{N} 
&=&\frac{d^k}{d\eta^k}\left.\frac{1}{2\pi  i}
\int_{\overline N- i\infty}^{\overline N+ i\infty} 
dNx^{-N}N^{-\eta}\right|_{\eta=0}
\nonumber \\
&=&\left.\frac{d^k}{d\eta^k} 
\bigg\{\Delta(\eta)\[\ln^{ \eta-1}\frac{1}{x}\]_+
\right|_{\eta=0 }+\delta(1-x)\bigg\}\\
&=&\[
\frac{1}{\ln\frac{1}{x}}\sum_{n=1}^{k}
\(\begin{array}{c}k\\n\end{array}\)
n\Delta^{(n-1)}(1)\(\ln\ln\frac{1}{x}\)^{k-n}\]_++\delta_{k0}\delta(1-x),\nonumber
\eea
where in the last step we used $\Delta^{(n)}(0)=n\Delta^{(n-1)}(1).$


\begin{thebibliography}{99}
\baselineskip14pt
\bibitem{cnt} 
S.~Catani and L.~Trentadue,
Nucl.\ Phys.\ B {\bf 327} (1989) 323.
\bibitem{sterman}
G.~Sterman,
Nucl.\ Phys.\ B {\bf 281} (1987) 310.
\bibitem{bassetto}
D.~Amati, A.~Bassetto, M.~Ciafaloni, G.~Marchesini and G.~Veneziano,
  Nucl.\ Phys.\ B {\bf 173} (1980) 429.
\bibitem{efres}
  A.~V.~Manohar,
  Phys.\ Rev.\  D {\bf 68} (2003) 114019;\\
  A.~Idilbi, X.~d.~Ji and F.~Yuan,
  Nucl.\ Phys.\  B {\bf 753} (2006) 42;\\
 T.~Becher, M.~Neubert and B.~D.~Pecjak,
  JHEP {\bf 0701}, 076 (2007).
\bibitem{fr} S.~Forte and G.~Ridolfi,
Nucl.\ Phys.\ B {\bf 650} (2003) 229.
\bibitem{cmnt}
  S.~Catani, M.~L.~Mangano, P.~Nason and L.~Trentadue,
  Nucl.\ Phys.\ B {\bf 478} (1996) 273;
\bibitem{frru}
  S.~Forte, G.~Ridolfi, J.~Rojo and M.~Ubiali,
  Phys.\ Lett.\  B {\bf 635} (2006) 313.
\bibitem{gardi}   E.~Gardi, G.~P.~Korchemsky, D.~A.~Ross and S.~Tafat,
  Nucl.\ Phys.\  B {\bf 636} (2002) 385.
\bibitem{lor}   T.~O.~Eynck, E.~Laenen and L.~Magnea,
  JHEP {\bf 0306} (2003) 057.
\bibitem{akh}  R.~Akhoury, M.~G.~Sotiropoulos and G.~Sterman,
  Phys.\ Rev.\ Lett.\  {\bf 81} (1998) 3819.
\bibitem{vogt}
 A.~Vogt,
  Phys.\ Lett.\  B {\bf 497} (2001) 228.
\bibitem{buras}
  W.~A.~Bardeen, A.~J.~Buras, D.~W.~Duke and T.~Muta,
  Phys.\ Rev.\  D {\bf 18} (1978) 3998.
\bibitem{stirling}
 A.~Kulesza and W.~J.~Stirling,
  JHEP {\bf 0312} (2003) 056.
\end{thebibliography}
\end{document}